\documentclass[12pt]{article}
\usepackage{amsmath,amssymb,bbm}
\usepackage{amsmath}
\begin{document}

\begin{center}

{\LARGE {A torsion-free background solution of the string theory}}

\end{center}

\vskip 0.7cm
\begin{center}
{\large Eun Kyung Park\footnote{E-mail: ekpark@pusan.ac.kr} and Pyung Seong Kwon\footnote{E-mail: bskwon@ks.ac.kr}\\
\vskip 0.2cm}
\end{center}

\vskip 0.05cm

\begin{center}
{\it $^1$Department of Physics, Pusan National University,\\ Busan 46241, Korea \\

\vskip 0.1cm

$^2$Department of Energy Science, Kyungsung University,\\ Busan 48434, Korea}
\end{center}

\thispagestyle{empty}

\vskip 2.0cm
\begin{center}
{\bf Abstract}
\vskip 0.1cm
\end{center}

\vskip 0.25cm
We find a background solution of the string theory which has a special property distinguished from the usual background solutions. This background solution does not produce the NS-NS two-form fields under T-duality and therefore the background vacua described by this solution essentially do not involve NS-NS type branes in their configurations, unlikely to the case of the ordinary Calabi-Yau ansatz. As a result the non-linear $\sigma$-models, whose target space metrics are given by these T-dual partners, can both be torsion-free.

\vskip 7.0cm
\medskip
\begin{center}
\end{center}

\newpage
\setcounter{page}{1}
\setcounter{footnote}{0}

\baselineskip 6.0mm

\vskip 1cm
\hspace{-0.65cm}{\bf \Large I. Introduction}
\vskip 0.5cm
\setcounter{equation}{0}
\renewcommand{\theequation}{1.\arabic{equation}}

In the usual compactifications of the 10D superstring theory the internal sector of the background vacuum is described by a Calabi-Yau manifold. Calabi-Yau manifolds generally contain singular points called conifold singularity (see \cite{1} for the reviews). So in the brane world models it is assumed that a stack of D3-branes, which is identified with our 3D external space, is located around \cite{2} or at certain conifold singularity \cite{3,8}.

Some years ago, however, it was argued \cite{4} that the background vacua of the string theory may involve NS5-branes implicitly in their Calabi-Yau or the conifold ansatz. The history of such an observation goes back to the paper on the $(p+3)$-dimensional string theory \cite{5}. In \cite{5}, it was shown that the existence of NS-NS type $p$-brane is essential to obtain flat geometries $R_2$ or $R_2 / {\mathcal Z}_n$ on the transverse directions and the usual codimensions-2 brane solutions with these background geometries already contain NS-NS type branes implicitly in their ansatz. Similar thing also happens in the codimensions-1 brane solutions of the five dimensional models compactified on $S_1 / {\mathcal Z}_2$. In \cite{6}, it was shown that in string theoretical setup the existence of the background NS-branes are indispensable to obtain flat geometry $M_4 \times S_1 / {\mathcal Z}_2$.

Such an observation is continued to the case of the 10D full-fledged string theory where the background geometry of the internal dimensions is described by a Calabi-Yau threefold with some conifold singularities attached on it as mentioned above. It was shown in \cite{7,8} that two intersecting NS5-branes can be thought of as a T-dual configuration of the conifold singularity. For instance, a IIA configuration of D4-brane suspended between two orthogonal NS5-branes corresponds to a IIB metric of D3-brane at a conifold singularity \cite{8}, which supports the above statement regarding T-duality of the conifold singularity. As a definite example of this T-duality, consider a conifold metric
\begin{equation}
ds^2_{\rm conifold} = dr^2 + r^2 d \Sigma^2_{1,1}\,\,,
\end{equation}
where
\begin{equation}
 d \Sigma^2_{1,1} = \frac{1}{9} \Big( d \psi + \sum_{i=1}^{2} \cos \theta_i d \phi_i \Big)^2 + \sum_{i=1}^{2}  \frac{1}{6} \Big( d\theta_i^2 + \sin^2 \theta_i d\phi_i^2 \Big) \,\,.
\end{equation}
Under T-duality along the isometry direction $\psi$, (1.1) turns into
\begin{equation}
ds^2_{\rm T-dual} = dr^2 + \frac{9}{r^2} d \psi^2 + r^2 \sum_{i=1}^{2}  \frac{1}{6} \Big( d\theta_i^2 + \sin^2 \theta_i d\phi_i^2 \Big) \,\,
\end{equation}
plus two-form fields $B_{\psi \phi_i}$ which are given by $ \cos \theta_i$.

The metric (1.3) is very singular. The scalar curvature calculated from (1.3) is given by ${\mathcal R}=16/r^2$, which goes to infinity as $r$ goes to zero. Also since $B_{\psi \phi_i}$ are NS-NS two-form fields, the T-dual configuration described by (1.3) necessarily contains two NS5-branes each of which becomes the source of $B_{\psi \phi_i}$. This suggests that the conifold metric, which is a Ricci-flat metric, is equivalent to a non-Ricci-flat singular metric plus NS5-branes described by $B_{\psi \phi_i}$. This example is very reminiscent of codimendion-1 \cite{6} and codimension-2 \cite{5} brane world models, where each of flat geometries of the transverse dimensions can be formally described by a sum of singular metric and NS-NS type branes. In this paper we want to check the possibility of having special solution distinguished from these background solutions described so far. Indeed, we show that there exists a background solution which has a unique property that it does not produce the NS-NS two-form fields under T-duality and therefore the NS-NS type branes are not involved unlikely to the case of the ordinary Calabi-Yau (or the conifold) ansatz.

\vskip 1cm
\hspace{-0.65cm}{\bf \Large II. Total action}
\vskip 0.5cm
\setcounter{equation}{0}
\renewcommand{\theequation}{2.\arabic{equation}}

The 10D action is given by
\begin{equation}
 I_{\rm bulk}= \frac{1}{2 \kappa_{10}^2} \int d^{10}x \sqrt{-G_{10}} \bigg( e^{-2 \Phi} \Big( {\mathcal R}_{10} + 4 (\nabla \Phi)^2  - \frac{1}{2
 \cdot 3!} H_{(3)}^2 \Big) - \frac{1}{2 \cdot 5!} {F}_{(5)}^2  \bigg) \,\,,
\end{equation}
where $H_{(3)}$($\equiv dB_{(2)}$) is the field strength of the NS-NS two-form $B_{(2)}$ and similarly $F_{(5)}$($\equiv dA_{(4)}$) the field strength of the R-R four-form $A_{(4)}$. We start with the configurations without NS5-branes, which is necessary to obtain Ricci-flat solutions. So we have $H_{(3)}=0$ in the action (2.1). Though we start with the case $H_{(3)}=0$ to obtain Ricci-flat solutions, it is always possible that those Ricci-flat solutions become nonsingular solutions with NS5-branes (i.e. with $H_{(3)} \neq 0$) under T-duality transformation as mentioned above.

Now we introduce an ansatz for the 10D metric as
\begin{equation}
ds_{10}^2 = e^{A({\hat r})} ds_{7}^{2} +  e^{B({\hat r})} d {\vec x}_3^2 \,\,,
\end{equation}
where $d {\vec x}_3^2$ represents the 3D external space $d {\vec x}_3^2 = dx_1^2 + dx_2^2 + dx_3^2$ and $ds_{7}^{2}$ is a 7D metric defined by
\begin{equation}
ds_{7}^{2} = -N^2 ({\hat r}) d{\hat t}^2 + \frac{d {\hat r}^2}{f^2 ({\hat r})} + \frac{U^2 ({\hat r})}{9} \Big( d \psi + \sum_{i=1}^{2} p_i \cos \theta_i d \phi_i \Big)^2 + \frac{V^2 ({\hat r})}{6} \sum_{i=1}^{2} q_i \Big( d\theta_i^2 + \sin^2 \theta_i d\phi_i^2 \Big) \,\,.
\end{equation}
Note that in the absence of the $d{\hat t}^2$ term the metric (2.3) becomes the conifold metric for $f({\hat r})=1$ and $U({\hat r})=V({\hat r})={\hat r}$. We similarly introduce an ansatz for $A_{(4)}$ as
\begin{equation}
A_{(4)} = \xi ({\hat r}) \,d{\hat t} \wedge dx_1 \wedge dx_2 \wedge dx_3 \,\,,
\end{equation}
and therefore
\begin{equation}
F_{(5)} = \partial_{\hat r} \xi ({\hat r}) \, d{\hat r} \wedge  d{\hat t} \wedge dx_1 \wedge dx_2 \wedge dx_3 \,\,.
\end{equation}

Now setting $e^{\Phi ({\hat r})} \equiv g_s  e^{{\Phi_s} ({\hat r})}$ and choosing $A({\hat r})$ as
\begin{equation}
A({\hat r}) = \frac{4}{5} {\Phi_s} ({\hat r}) - \frac{3}{5} B({\hat r}) \,\,,
\end{equation}
one finds that $I_{\rm bulk}$ in (2.1) reduces to a 7D action given by\footnote{In (2.7) the squares of partial derivatives of the form $(\partial A)(\partial B)$ represent $(\partial A)(\partial B) \equiv g^{AB} (\partial_A A)(\partial_B B)$, where $g^{AB}$ is the inverse of the 7D metric $g_{AB}$.}
\begin{equation}
\frac{I_{\rm bulk}}{ \big(\int d^3 {\vec x}\, \big)} =  \frac{1}{2 \kappa_{10}^2 g_s^2} \int d^{7}x \, \sqrt{- {g}_{_{_7}}} \Big ( {\mathcal R}_{7} - \frac{4}{5} (\partial {\Phi_s})^2  + \frac{6}{5} (\partial {\Phi_s}) (\partial B) - \frac{6}{5} (\partial B)^2 \nonumber
\end{equation}
\begin{equation}
+ \frac{g_s^2}{2} \frac{e^{M}}{N^2} (\partial \xi)^2 \Big) \,\,,
\end{equation}
where $M \equiv \frac{6}{5} \Phi_s - \frac{12}{5}B$, and ${\mathcal R}_{7} / \sqrt{- {g}_{_{_7}}}$ are the Ricci scalar/determinant of the 7D metric (2.3), respectively.

In addition to this, we also have an action for the D3-branes which are R-R sources of $A_{(4)}$ :
\begin{equation}
I_{\rm brane} = - \int d^{3+1} x \,\sqrt{-det |G_{\mu\nu}|} \, T(\Phi) + \mu_3 \int A_{(4)} \,\,,
\end{equation}
where $G_{\mu\nu}$ is a pullback of the 10D metric $G_{MN}$ to the 4D brane world and $T(\Phi)$ is the D3-brane tension given by $T(\Phi)=T_{3} e^{- \Phi}$ at the tree level. As in (2.7) the brane action (2.8) also reduces to a 7D action as
\begin{equation}
\frac{I_{\rm brane}}{\big(\int d^3 {\vec x} \,\big)} = - \int d^{7}x \, \sqrt{- {g}_{_{_7}}} \, e^{\frac{2}{5} {\Phi_s} + \frac{6}{5} B} T(\Phi) \, \delta^6 ({\vec{\hat r}}\,) + \mu_3 \int d^{7}x \, \sqrt{h_6} \, \xi ({\hat r}) \, \delta^6 ({\vec{\hat r}}\,) \,\,,
\end{equation}
where $h_6$ is the determinant of the 6D metric $ds_{6}^2$ ($\equiv ds_{7}^2 + N^2 ({\hat r}) d{\hat t}^2$) and the delta function $\delta^6 ({\vec{\hat r}}\,)$ is defined by $\int d^{6}y \, \sqrt{h_6} \, \delta^6 ({\vec{\hat r}}\,) =1$. The total 7D action is now given by the sum of (2.7) and (2.9) :
\begin{equation}
I_{\rm total}^{7D} \equiv \frac{\big( I_{\rm bulk} + I_{\rm brane} \big)} {\big(\int d^3 {\vec x} \,\big)} \,\,.
\end{equation}

\vskip 1cm
\hspace{-0.65cm}{\bf \Large III. 7D field equations}
\vskip 0.5cm
\setcounter{equation}{0}
\renewcommand{\theequation}{3.\arabic{equation}}

\vskip 0.3cm
\hspace{-0.6cm}{\bf \large 3.1 Equation for $\xi (\hat{r})$}
\vskip 0.15cm

Since we have $I_{\rm total}^{(7D)}$, we can now find field equations for the ${\hat r}$-dependent fields defined on the internal dimensions from this action. Let us start with the field equation for $\xi ({\hat r})$ first. From (2.10) we obtain
\begin{equation}
\frac{1}{\sqrt{h_6}} \partial_m \Big( \frac{e^M}{N} \,\sqrt{h_6} h^{mn} (\partial_n \xi) \Big) =  2 \kappa_{10}^2 \mu_3  \delta^6 ({\vec{\hat r}}\,) \,\,,
\end{equation}
where $h_{mn}$ represents the 6D metric $ds_6^2 \equiv h_{mn} dy^m dy^n$, so we have $\sqrt{- {g}_{_{_7}}} = N \sqrt{h_6}$ in (2.7) and (2.9). Upon integration $\int d^6 y \sqrt{h_6}$, (3.1) gives
\begin{equation}
\partial_{\hat r} \xi = \frac{ 2 \kappa_{10}^2 \mu_3 }{\rm Vol(B)}\, e^{-M}\, \frac{N}{f} \,\frac{1}{UV^4} \,\,,
\end{equation}
where $\rm{Vol(B)}$ is the volume of the base of the cone with unit radius : ${\rm Vol(B)} = \int \epsilon_5$ with $\epsilon_5 = \sqrt{det |{\hat h}_{ab}|} \,d \psi \wedge d \theta_1  \wedge d \phi_1  \wedge d \theta_2  \wedge d \phi_2$, and where ${\hat h}_{ab}$ is defined by
\begin{equation}
d \Sigma_{p_{_1}, \, p_{_2}}^2 = \frac{1}{9} \Big( d \psi + \sum_{i=1}^{2} p_i \cos \theta_i d \phi_i \Big)^2 + \frac{1}{6} \sum_{i=1}^{2} q_i \Big( d\theta_i^2 + \sin^2 \theta_i d\phi_i^2 \Big) \equiv {\hat h}_{ab} d {y}^a d {y}^b \,\,.
\end{equation}
From (2.5) and (3.2) one finds
\begin{equation}
\int {}^* F_{(5)} =  2 \kappa_{10}^2 \mu_3 \,\,,
\end{equation}
which shows that $\mu_3$ is an R-R electric charge carried by a D3-brane and this confirms the fact that (3.2) is correct.

\vskip 0.3cm
\hspace{-0.6cm}{\bf \large 3.2 Field equations in covariant forms}
\vskip 0.15cm

Now we turn to the Einstein equation and equations for $\Phi_s$ and $B$. In the covariant form the 7D Einstein equation can be rewritten as
\begin{equation}
{\mathcal R}_{AB} - \frac{1}{2} {g}_{AB} {\mathcal R}_7 = \kappa_{10}^2 (T_{AB}+t_{AB})
\end{equation}
with
\begin{equation}
T_{AB} = \frac{4}{5\kappa_{10}^2} \big[ (\partial_A \Phi_s )(\partial_B \Phi_s )-\frac{1}{2} {g}_{AB}(\partial \Phi_s )^2 \big] -  \frac{6}{5\kappa_{10}^2} \big[ (\partial_A \Phi_s )(\partial_B B )-\frac{1}{2} {g}_{AB}(\partial \Phi_s ) (\partial B ) \big] \nonumber
\end{equation}
\begin{equation}
 + \frac{6}{5\kappa_{10}^2} \big[ (\partial_A B )(\partial_B B )-\frac{1}{2} {g}_{AB}(\partial B)^2 \big]~~~~~~~~~~~~~~~~~~~~~~~~~~~~~~~~~~~~~~~~~~~~~~~~~~ \nonumber
\end{equation}
\begin{equation}
+\frac{1}{2\kappa_{10}^2}\, e^M g^{00}  \big[ (\partial_A \xi )(\partial_B \xi )-\frac{1}{2} {g}_{AB}(\partial \xi)^2 + g_{00} \delta_A^0 \delta_B^0 (\partial \xi)^2 \big]\,\,, ~~~~~~~~~~~~~~~~~
\end{equation}
and
\begin{equation}
t_{AB}=- g_s^2 \, g_{0A} {g}^{00} {g}_{0B}\, e^{\frac{2}{5} {\Phi_s} + \frac{6}{5} B} T(\Phi) \, \delta^6 ({\vec{\hat r}}\,)\,\,,~~~~~~~~~~~~~~~~~~~~~~~~~~~~~~~~~~~~~~~~~~~~~~~~~~~~~
\end{equation}
where $(A,B)$ are 7D indices and $g_{00}$ represents the $tt$-component of $g_{AB}$ etc. Also, the equations for $\Phi_s$ and $B$ become
\begin{equation}
\frac{8}{5} \frac{1}{\sqrt{- {g}_{_{_7}}}} \partial_A \big({\sqrt{- {g}_{_{_7}}}} g^{AB} \partial_B \Phi_s \big) - \frac{6}{5} \frac{1}{\sqrt{- {g}_{_{_7}}}} \partial_A \big({\sqrt{- {g}_{_{_7}}}} g^{AB} \partial_B B \big) ~~~~~~~~~~~~~~~~~~~~~~~~~\nonumber
\end{equation}
\begin{equation}
-\frac{3}{5} e^M g^{00} (\partial \xi)^2 =  \frac{4}{5} \kappa_{10}^2 g_s^2 e^{\frac{2}{5} {\Phi_s} + \frac{6}{5} B} \Big( T (\Phi) + \frac{5}{2} \frac{\partial  T (\Phi)}{\partial \Phi} \Big) \delta^6 ({\vec{\hat r}}\,) \,\,,
\end{equation}
\begin{equation}
\frac{12}{5} \frac{1}{\sqrt{- {g}_{_{_7}}}} \partial_A \big({\sqrt{- {g}_{_{_7}}}} g^{AB} \partial_B B \big) - \frac{6}{5} \frac{1}{\sqrt{- {g}_{_{_7}}}} \partial_A \big({\sqrt{- {g}_{_{_7}}}} g^{AB} \partial_B \Phi_s \big) ~~~~~~~~~~~~~~~~~~~~~~~~~\nonumber
\end{equation}
\begin{equation}
+\frac{6}{5} e^M g^{00} (\partial \xi)^2 =  \frac{12}{5} \kappa_{10}^2 g_s^2  e^{\frac{2}{5} {\Phi_s} + \frac{6}{5} B}  T (\Phi) \delta^6 ({\vec{\hat r}}\,) \,\,.
\end{equation}

\vskip 0.3cm
\hspace{-0.6cm}{\bf \large 3.3 Field equations in component forms}
\vskip 0.15cm

Now we put
\begin{equation}
U (\hat{r}) = \gamma (\hat{r}) R (\hat{r})\,\,, ~~~~~V(\hat{r}) = R(\hat{r})\,\,,
\end{equation}
in the metric (2.3): i.e. we consider the 7D metric of the form
\begin{equation}
ds_7^2 = - N^2 (\hat{r}) d{\hat t}^2 + \frac{d\hat{r}^2}{f^2 (\hat{r})} + \frac{\gamma^2 (\hat{r}) R^2 (\hat{r})}{9} \Big( d \psi + \sum_{i=1}^{2} p_i \cos \theta_i d \phi_i \Big)^2 ~~~~~~~\nonumber
\end{equation}
\begin{equation}
+ \frac{R^2 (\hat{r})}{6} \sum_{i=1}^{2} q_i \Big( d\theta_i^2 + \sin^2 \theta_i d\phi_i^2 \Big)\,\,.
\end{equation}
For the given metric (3.11) the independent components of the Einstein equation (3.5) take the forms
\begin{equation}
 N(fR^{\prime})^{\prime} + \frac{1}{25}\, NfR \, \mathcal{H} + \frac{g_s^2}{20}\, \frac{fR}{N}\, e^M {\xi^{\prime}}^2 + 2 NfR \, \Big( \frac{{R^{\prime}}^2}{R^2} - \frac{\alpha}{f^2 R^2} \Big)  + \frac{1}{5} \, NfR \, \frac{(f \gamma^{\prime})^{\prime} }{f \gamma}  \nonumber
\end{equation}
\begin{equation}
 + \frac{6}{5}\,NfR\, \frac{{R^{\prime}}}{R} \frac{{\gamma^{\prime}}}{\gamma}  \,=\, - \frac{\kappa^2_{10}  g_s^2 }{5} \,\frac{NR}{f}\, {e^{\frac{2}{5} {\Phi_s} + \frac{6}{5} B}} \, T (\Phi) \delta^6 (\vec{\hat{r}}\,) \,\,,
\end{equation}
\begin{equation}
 N^{\prime}f R^{\prime} - \frac{1}{25}\,NfR\, \mathcal{H} + \frac{g_s^2}{20} \, \frac{fR}{N}\, e^M {\xi^{\prime}}^2 + 2 NfR \, \Big( \frac{{R^{\prime}}^2}{R^2} - \frac{\alpha}{f^2 R^2} \Big)  \nonumber
\end{equation}
\begin{equation}
+ \frac{1}{5} \,NfR\, \Big( \frac{{N^{\prime}}}{N}+ 4\frac{{R^{\prime}}}{R}\Big) \frac{{\gamma^{\prime}}}{\gamma} \,=\,0 \,\,,
\end{equation}
\begin{equation}
 (N^{\prime} f)^{\prime} R + 4 \Big[N(fR^{\prime})^{\prime} + N^{\prime}f R^{\prime} \Big]+ \frac{1}{5}\,NfR\, \mathcal{H} - \frac{g_s^2}{4} \, \frac{fR}{N}\, e^M {\xi^{\prime}}^2 + 6 \,NfR\,\Big( \frac{{R^{\prime}}^2}{R^2} - \frac{\alpha_{ij}}{f^2 R^2} \, \Big) \nonumber
\end{equation}
\begin{equation}
~~~~~~+ NfR\,\frac{(f \gamma^{\prime})^{\prime}}{f \gamma} + NfR\,\Big( \frac{{N^{\prime}}}{N} + 5\frac{{R^{\prime}}}{R}\Big) \frac{{\gamma^{\prime}}}{\gamma} \,=\,0 \,\,,~~~(i,j=1,2~\,{\rm and}~\,i \neq j)\,,
\end{equation}
\begin{equation}
 (N^{\prime} f)^{\prime} R + 4 \Big(N(fR^{\prime})^{\prime} + N^{\prime}f R^{\prime} \Big)+ \frac{1}{5}\,NfR\, \mathcal{H} - \frac{g_s^2}{4} \, \frac{fR}{N}\, e^M {\xi^{\prime}}^2 + 6 \,NfR\,\Big( \frac{{R^{\prime}}^2}{R^2} - \frac{\alpha}{f^2 R^2} \, \Big) \nonumber
\end{equation}
\begin{equation}
+2 \,NfR \, \Big(\sum_{i=1}^{2} \Big(\frac{p_i}{q_i}\Big)^2 \gamma^2 -2 \alpha \Big) \frac{1}{f^2 R^2} \,=\,0 \,,
\end{equation}
where $\mathcal{H} \equiv 2 {\Phi_s^{\prime}}^2 - 3\Phi_s^{\prime} B^{\prime} + 3{B^{\prime}}^2$ and the "prime" denotes the derivative with respect to $\hat{r}$. Also $\alpha$ and $\alpha_{ij}$ are defined by
\begin{equation}
\alpha = - \frac{1}{10} \Big(\sum_{i=1}^{2} \Big(\frac{p_i}{q_i}\Big)^2 \gamma^2 - \sum_{i=1}^{2} \frac{6}{q_i} \,\Big)\,,
\end{equation}
\begin{equation}
\alpha_{ij} = \frac{1}{6} c_{ij} \gamma^2 + \frac{1}{q_j} \,,
\end{equation}
\begin{equation}
c_{ij} \equiv \Big(\frac{p_i}{q_i}\Big)^2 - \Big(\frac{p_j}{q_j}\Big)^2 ~\,{\rm with}~\,i \neq j \,\,.
\end{equation}

Besides these, we also have the equations for $\Phi_s$ and $B$. The linear combinations of those equations give
\begin{equation}
(Nf \gamma R^5 \Phi_s^{\prime})^{\prime}  \,=\,0\,\,,
\end{equation}
\begin{equation}
(Nf \gamma R^5 B^{\prime})^{\prime} - \frac{1}{2} \,\frac{f}{N}\,\gamma R^5\, {e^M}\,{\xi^{\prime}}^2 \,=\, \kappa^2_{10} g_s^2 \,\frac{N}{f}\,\gamma R^5\, e^{\frac{2}{5} {\Phi_s} + \frac{6}{5} B} \, T (\Phi)  \delta^6 (\vec{\hat{r}}\,) \,\,,
\end{equation}
where we have used the fact that $T(\Phi)$ is given by $T(\Phi) = T_3 e^{-\Phi}$ at the tree level and therefore we have $\frac{\partial  T (\Phi)}{\partial \Phi} = - T (\Phi)$. Also in the whole above equations ${\xi^{\prime}}^2$ terms can be rewritten (see (3.2)) as
\begin{equation}
\frac{f}{N}\,\gamma R^5\, {e^M}\,{\xi^{\prime}}^2 \,=\, \frac{N}{f}\,\frac{1}{\gamma R^5} \,e^{-M} \frac{Q_D^2}{g_s^2} \,\,,~~~~~~\Big(Q_D \equiv \frac{2 \kappa^2_{10} g_s \mu_3}{\rm Vol(B)}\,\Big)\,\,.
\end{equation}

\vskip 1cm
\hspace{-0.65cm}{\bf \Large IV. Solutions to the field equations in the absence of D3-brane}
\vskip 0.5cm
\setcounter{equation}{0}
\renewcommand{\theequation}{4.\arabic{equation}}

In the previous section we found a set of field equations defined on the internal dimensions. This set of field equations basically consists of four Einstein equations (eqs. (3.12) to (3.15)) plus equations for $\Phi_s$ and $B$ (eqs. (3.19) and (3.20)). In this section we want to solve these equations to obtain some nontrivial solutions which do not involve NS-NS type branes in both of before and after T-duality transformation. Before we start our discussion we have to notice that the equation (3.14) must be satisfied for both $\alpha_{12}$ and $\alpha_{21}$. So we must require $\alpha_{12} = \alpha_{21}$, which then implies
\begin{equation}
c_{12} \gamma^2 ({\hat r}) - 3 \big( \frac{1}{q_1} - \frac{1}{q_2} \big) =0
\end{equation}
from (3.17). Equation (4.1) imposes a strong constraint on $p_i$ and $q_i$, especially when $\gamma ({\hat r})$ is a nontrivial function of ${\hat r}$.

Now we start the procedure to obtain the solutions of the field equations. In this section we first consider the cases in which the D3-branes are absent. In these cases we basically obtain Ricci-flat solutions by the following reasons. In the absence of D3-branes ($T_3 = Q_D =0$), the field equations (3.19) and (3.20) reduce to
\begin{equation}
(Nf \gamma R^5 \Phi^{\prime}_s )^{\prime} =0 ~~~~~~~{\rm and}~~~~~~~ (Nf \gamma R^5 B^{\prime})^{\prime} =0 \,\,,
\end{equation}
and therefore we have
\begin{equation}
\Phi_s ({\hat r}) = B ({\hat r}) =0 \,\,.
\end{equation}
Also since $A_{(4)}$ (and therefore $\xi ({\hat r})$) vanishes in the absence of D3-branes, $T_{AB}$ in (3.6) vanishes by (4.3). Finally $t_{AB}$ in (3.7) also vanishes by $T_3 =0$, and therefore the Einstein equations (3.5) reduces to
\begin{equation}
 {\mathcal R}_{AB} =0
\end{equation}
in the case with no D3-branes. So the solutions of this section are necessarily Ricci-flat in the 7D space defined by (2.3).

To obtain the solutions to the field equations we take an ansatz for $N(\hat{r})$ and $R({\hat r})$ as
\begin{equation}
N (\hat{r}) = f (\hat{r})\,\,,~~~~~ R({\hat r}) = {\hat r} \,\,.
\end{equation}
Since $\mathcal{H} = {\xi^{\prime}}^2 =0$ by $T_3 =Q_D =0$ (see (3.21) and (4.3)), the Einstein equations (3.12) to (3.15) now reduce to\footnote{Note that the equations (4.6) to (4.9) were obtained from (3.5), not from (4.4). But the solutions of (4.6) to (4.9) also satisfy (4.4) as described above.}
\begin{equation}
f f^{\prime} + 2(f^2 - \alpha)\, \frac{1}{\hat r} + \frac{1}{5} \frac{f(f \gamma^{\prime})^{\prime} }{\gamma} \,{\hat r} + \frac{6}{5} \, f^2 \, \frac{{\gamma^{\prime}}}{\gamma} =0 \,\,,
\end{equation}
\begin{equation}
f f^{\prime} + 2(f^2 - \alpha)\, \frac{1}{\hat r} + \frac{1}{5} f f^{\prime} \frac{\gamma^{\prime}}{\gamma}\, {\hat r} + \frac{4}{5} \, f^2 \, \frac{{\gamma^{\prime}}}{\gamma} =0 \,\,,
\end{equation}
\begin{equation}
\big(f f^{\prime} \big)^{\prime} + 8 f f^{\prime} \, \frac{1}{\hat r} + 6(f^2 - \alpha_{ij}) \,\frac{1}{{\hat r}^2} + \frac{f(f \gamma^{\prime})^{\prime} }{\gamma} + f f^{\prime} \, \frac{\gamma^{\prime}}{\gamma} + \frac{5}{\hat r} \, f^2 \,\frac{{\gamma^{\prime}}}{\gamma} =0 \,\,,
\end{equation}
\begin{equation}
\big(f f^{\prime} \big)^{\prime} + 8 f f^{\prime} \, \frac{1}{\hat r} + 6(f^2 - \alpha) \,\frac{1}{{\hat r}^2} + 2 \Big(\sum_{i=1}^{2} \Big(\frac{p_i}{q_i}\Big)^2 \gamma^2 -2 \alpha \Big)\, \frac{1}{{\hat r}^2} =0 \,.
\end{equation}
Then from (4.6) and (4.7) we obtain
\begin{equation}
\gamma^{\prime \prime} + \frac{2}{\hat r} \, \gamma^{\prime} =0 \,\,,
\end{equation}
and this equation is solved by
\begin{equation}
\gamma ({\hat r}) = a + \frac{b}{\hat r} \,\,.
\end{equation}
Similarly from (4.8) and (4.9), and using (4.10) we obtain
\begin{equation}
\Big( 2 \, f {f^{\prime}} + \frac{3}{\hat r} f^2 \Big) \frac{ \gamma^{\prime}}{\gamma} +  \Big( 10 \alpha - 6 \alpha_{ij} - 2 \sum_{i=1}^{2} \Big(\frac{p_i}{q_i}\Big)^2 \gamma^2 \Big) \frac{1}{{\hat r}^2} =0\,\,,
\end{equation}
which acts as a constraint on $p_i$ and $q_i$ especially when $\gamma ({\hat r})$ is constant. In the next we solve the above equations for the cases where $\gamma ({\hat r}) ={\rm constant}$ ($b=0$), or $\gamma ({\hat r}) = a + \frac{b}{\hat r} $ with $b \neq 0$.

\vskip 0.3cm
\hspace{-0.6cm}{\bf \large 4.1 Case I: $\gamma ({\hat r}) ={\rm constant} \equiv \gamma_0$ }($b=0$)
\vskip 0.15cm

In this case the constraint equation (4.1) reduces to
\begin{equation}
c_{12} \gamma^2_0 - 3 \big( \frac{1}{q_1} - \frac{1}{q_2} \big) =0 \,\,,
\end{equation}
and in particular one can show that (4.13) can be solved by
\begin{equation}
{\rm (a)~Case~A~}:~ \gamma_0 =1\,\,,~~p_1 =p_2 =1\,\,, ~~q_1 =q_2 =1 \,\,,~~~~~~~~~~~~~~~~~~~~~~~~~~~~~~
\end{equation}
and
\begin{equation}
{\rm (b)~Case~B~}:~ \gamma_0 = \frac{3}{2 \sqrt{2}}\,\,,~~p_1 =1\,\,, p_2 =0\,\,, ~~q_1 = \frac{3}{4}\,\,, q_2 = \frac{3}{2} \,\,.~~~~~~~~~~~~~~~~~
\end{equation}
Indeed, these are two well-known special cases of the conifold metric in which the 5D bases are given respectively by
\begin{equation}
{\rm (a)~} d \Sigma_{1,1}^2 = \frac{1}{9} \Big( d \psi + \cos \theta_1 d \phi_1 + \cos \theta_2 d \phi_2 \Big)^2 + \frac{1}{6} \Big( d\theta_1^2 +
\sin^2 \theta_1 d\phi_1^2 \Big) + \frac{1}{6} \Big( d\theta_2^2 + \sin^2 \theta_2 d\phi_2^2 \Big) \,\,,
\end{equation}
and
\begin{equation}
{\rm (b)~} d \Sigma_{1,0}^2 = \frac{1}{8} \Big( d \psi + \cos \theta_1 d \phi_1 \Big)^2 + \frac{1}{8} \Big( d \theta_1^2 + \sin^2 \theta_1 d \phi_1^2 \Big) + \frac{1}{4} \Big( d \theta_2^2 + \sin^2 \theta_2 d \phi_2^2 \Big) \,\,.~~~~~~~~~~
\end{equation}

In addition to (4.13), we also have the independent equations following from (4.6) to (4.9) :
\begin{equation}
f f^{\prime} + 2(f^2 - \alpha)\, \frac{1}{\hat r} = 0 \,\,,
\end{equation}
\begin{equation}
\big(f f^{\prime} \big)^{\prime} + 8 f f^{\prime} \, \frac{1}{\hat r} + 6(f^2 - \alpha_{ij}) \,\frac{1}{{\hat r}^2} = 0 \,\,,
\end{equation}
\begin{equation}
10 \alpha - 6 \alpha_{ij} - 2  \sum_{i=1}^{2} \Big(\frac{p_i}{q_i}\Big)^2 \gamma_0^2 =0 \,\,,
\end{equation}
where the last equation (4.20) is consistent with (4.12) because $\gamma ({\hat r}) ={\rm constant}$ in the case I. (4.20) constitutes together with (4.13) a set of constraint equations for $p_i$ and $q_i$ of the case I and one can check that the constants in (4.14) and (4.15) really satisfy both (4.13) and (4.20).

Using (3.16) and (3.17), and also using $c_{12} = \big(\frac{p_1}{q_1}\big)^2 - \big(\frac{p_2}{q_2}\big)^2 = - c_{21}$, one finds that the two independent components of the equation (4.20) reduce to
\begin{equation}
\Big( 2 \Big(\frac{p_1}{q_1}\Big)^2 + \Big(\frac{p_2}{q_2}\Big)^2 \Big) \gamma_0^2 - \frac{3}{q_1} =0 \,\,,
\end{equation}
\begin{equation}
\Big( \Big(\frac{p_1}{q_1}\Big)^2 + 2 \Big(\frac{p_2}{q_2}\Big)^2 \Big) \gamma_0^2 - \frac{3}{q_2} =0 \,\,,
\end{equation}
and one can check that these two equations are consistent with (4.13). Now we add (4.21) and (4.22) to obtain a constraint equation
\begin{equation}
\Big( \Big(\frac{p_1}{q_1}\Big)^2 + \Big(\frac{p_2}{q_2}\Big)^2 \Big) \gamma_0^2 = \frac{1}{q_1} + \frac{1}{q_2} \,\,.
\end{equation}
By this constraint equation we find from (3.16) that $\alpha$ is given by
\begin{equation}
\alpha = \frac {1}{2} \Big( \frac{1}{q_1} + \frac{1}{q_2} \Big) \,\,,
\end{equation}
and this $\alpha$ must be positive by (4.23) again. Besides this, $\alpha_{12}$ in (3.17) also reduces by (4.13) into
\begin{equation}
\alpha_{12} = \frac {1}{2} \Big( \frac{1}{q_1} + \frac{1}{q_2} \Big) = \alpha_{21} \,\,,
\end{equation}
so we have
\begin{equation}
\alpha = \alpha_{ij} \geq 0 \,\,.
\end{equation}
Finally from (4.13) and (4.23) we obtain
\begin{equation}
\frac{2}{q_1} - \frac{1}{q_2} = \Big(\frac{p_1}{q_1}\Big)^2  \gamma_0^2 ~~\rightarrow~~ \frac{q_2}{q_1} \geq \frac{1}{2} \,\,,
\end{equation}
\begin{equation}
\frac{2}{q_2} - \frac{1}{q_1} = \Big(\frac{p_2}{q_2}\Big)^2  \gamma_0^2 ~~\rightarrow~~ \frac{q_1}{q_2} \geq \frac{1}{2} \,\,,
\end{equation}
which gives a constraint on $q_i$ as
\begin{equation}
\frac{1}{2} \leq \frac{q_2}{q_1} \leq 2 \,\,.
\end{equation}

Turning back to the differential equations, using (4.26)  one can rewrite (4.18) and (4.19) as
\begin{equation}
X^{\prime} + \frac{4}{\hat r} X =0 \,\,,
\end{equation}
\begin{equation}
X^{\prime \prime} + \frac{8}{\hat r} X^{\prime} + \frac{12}{{\hat r}^2} X   =0 \,\,,
\end{equation}
where $X({\hat r})$ is defined by
\begin{equation}
X({\hat r}) \equiv f^2 ({\hat r}) - \alpha \,\,.
\end{equation}
Now we have to find $f({\hat r})$ which satisfies both (4.30) and (4.31). But these two equations are not independent of each other. (4.31) can be rewritten as
\begin{equation}
\Big( X^{\prime} + \frac{4}{\hat r} X \Big)^{\prime} + \frac{4}{\hat r} \Big( X^{\prime} + \frac{4}{\hat r} X \Big) =0 \,\,,
\end{equation}
and therefore the solution to (4.30) is also the solution of (4.31). The general solution of (4.30) is given by
\begin{equation}
 f^2 ({\hat r}) = \alpha - \frac{{\hat a}^4}{{\hat r}^4} \,\,,
\end{equation}
where ${\hat a}$ is an arbitrary constant with length dimension one. In this section we are essentially interested in the cases $A$ and $B$ in (4.14) and (4.15). In these cases $\alpha$ and $\alpha_{ij}$ are given by $\alpha = \alpha_{ij} =1$ and therefore $ f^2 ({\hat r})$ becomes
\begin{equation}
f^2 ({\hat r}) = 1 - \frac{{\hat a}^4}{{\hat r}^4} \,\,.
\end{equation}
Hence the 7D metric in (3.11) reduces to
\begin{equation}
ds_7^2 = - \Big( 1- \frac{{\hat a}^4}{{\hat r}^4} \Big) d{\hat t}^2 + \frac{d {\hat r}^2}{\Big( 1- \frac{{\hat a}^4}{{\hat r}^4} \Big)} + {\hat r}^2 \, d \Sigma_{p_{1}, \, p_{2}}^2 \,\,,
\end{equation}
which may be regarded as a black hole-type solution when ${\hat a} \neq 0$.

\vskip 0.3cm
\hspace{-0.6cm}{\bf \large 4.2 Case II: $\gamma ({\hat r}) = a + \frac{b}{\hat r} $ ($b \neq 0$)}
\vskip 0.15cm

Now we turn to the case where $\gamma ({\hat r})$ is given by (4.11) with $b \neq 0$. In this case (4.1) requires
\begin{equation}
q_1 = q_2 \equiv q \,\,,~~~ c_{12} =0 \,\,\rightarrow\,\, p_1 =p_2 \equiv p \,\,,
\end{equation}
and therefore $\alpha$ and $\alpha_{ij}$ become
\begin{equation}
\alpha = - \frac{1}{5} \Big(\, \frac{p^2}{q^2} \gamma^2 ({\hat r}) - \frac{6}{q} \,\Big)\,\,, ~~~~~~~\alpha_{12}=\alpha_{21} = \frac{1}{q} \,\,,
\end{equation}
and the remaining constraint (4.12) reduces to
\begin{equation}
2 ff^{\prime} + \frac{3}{\hat r} f^2 + 6 \frac{\gamma}{\gamma^{\prime}} \Big( \frac{1}{q} - \frac{p^2}{q^2} \gamma^2  \Big) \frac{1}{{\hat r}^2} =0\,\,.
\end{equation}
Since $\gamma = a+ \frac{b}{\hat r}$, (4.39) can be rewritten as
\begin{equation}
(f^2 )^{\prime} + \frac{3}{\hat r} f^2 = Q({\hat r}) \,\,,
\end{equation}
where
\begin{equation}
Q({\hat r}) = C_0 + \frac{C_1}{\hat r} + \frac{C_2}{{\hat r}^2}+  \frac{C_3}{{\hat r}^3} \,\,
\end{equation}
with
\begin{equation}
C_0 \equiv 6 \frac{a}{b} \Big( \frac{1}{q} - a^2 \frac{p^2}{q^2}\Big)  \,\,,~~~~~~~C_1 \equiv 6  \Big( \frac{1}{q} - 3 a^2 \frac{p^2}{q^2}\Big) \,\,,\nonumber
\end{equation}
\begin{equation}
C_2 \equiv - 18 ab \frac{p^2}{q^2} \,\,,~~~~~~~C_3 \equiv - 6 b^2 \frac{p^2}{q^2} \,\,.
\end{equation}
The general solution of (4.40) is given by
\begin{equation}
f^2 ({\hat r}) = \frac{C_0}{4} {\hat r} +  \frac{C_1}{3} +  \frac{C_2}{2} \frac{1}{\hat r}+ \frac{C_3}{{\hat r}^2}+  \frac{C}{{\hat r}^3} \,\,,
\end{equation}
where $C$ is an arbitrary constant with length dimension three.

Besides (4.39), the solution (4.43) must also satisfy the remaining equations
\begin{equation}
\frac{\gamma}{\gamma^{\prime}} \Big( 5 (f^2 )^{\prime} + \frac{20}{\hat r} \big( f^2 - \alpha \big) \Big) + {\hat r} (f^2 )^{\prime} +8f^2 =0 \,\,,
\end{equation}
\begin{equation}
\big(f^{2} \big)^{\prime \prime} + \frac{8}{\hat r} \big(f^{2} \big)^{\prime} + \frac{12}{{\hat r}^2} (f^2 - \alpha) +8 \Big( \frac{p^2}{q^2} \gamma^2 - \alpha \Big)\frac{1}{{\hat r}^2} =0 \,\,,
\end{equation}
which are rewrites of (4.7) and (4.9). Namely, together with (4.10) and (4.39), (4.44) and (4.45) constitute a new set of independent equations which substitutes for (4.6), (4.7), (4.8) and (4.9). Now substituting (4.43) into (4.44) gives constraints on $a$ and $p$ :
\begin{equation}
 \frac{a^2}{b^2} \Big( \frac{1}{q} - a^2 \, \frac{p^2}{q^2}\Big) =0  \,\,, ~~~~~ \frac{a}{b} \Big( \frac{2}{q} - 7 a^2 \, \frac{p^2}{q^2}\Big) =0 \,\,,  ~~~~~ a^2 \,\frac{p^2}{q^2} =0  \,\,,   \nonumber
\end{equation}
\begin{equation}
ab \, \frac{p^2}{q^2} =0  \,\,, ~~~~~  4 b^2 \, \frac{p^2}{q^2} - \frac{a}{b} C=0 \,\,,
\end{equation}
where $q$ and $b$ are nonzero constants. (4.46) requires
\begin{equation}
a=0\,\,, ~ p=0\,\,~\,\rightarrow ~\, C_0 = C_2 = C_3 =0\,\,, ~C_1 = \frac{6}{q}\,\,.
\end{equation}
So $\gamma ({\hat r})$ and $f^2 ({\hat r})$ in (4.43) reduce, respectively, to
\begin{equation}
\gamma ({\hat r}) = \frac{b}{\hat r}\,\,,~~~~~  f^2 ({\hat r}) = \frac{2}{q} +  \frac{C}{{\hat r}^3} \,\,,
\end{equation}
and one check that the set of functions in (4.48) also satisfies (4.45). After all this, we find that the 7D metric (3.11) finally becomes
\begin{equation}
ds_7^2 = - \frac{2}{q} \Big( 1- \frac{{\hat a}^3}{{\hat r}^3} \Big) d{\hat t}^2 + \frac{d {\hat r}^2}{\frac{2}{q}\Big( 1- \frac{{\hat a}^3}{{\hat r}^3} \Big)} + \frac{{\hat r}^2}{3} \sum_{i=1}^{2} \Big( d\theta_i^2 + \sin^2 \theta_i d\phi_i^2 \Big) + {\hat R}_0^2 d \psi^2\,\,,
\end{equation}
where ${\hat a}$ is an arbitrary constant and ${\hat R}_0 \equiv \frac{b}{3}$.

\vskip 1cm
\hspace{-0.65cm}{\bf \Large V. In the presence of D3-branes}
\vskip 0.5cm
\setcounter{equation}{0}
\renewcommand{\theequation}{5.\arabic{equation}}

So far we have considered the case $T_3 =Q_D =0$. In this section we turn to the case with nonzero $T_3$ and $Q_D$, which means that we now have a stack of D3-branes at the singularity ${\hat r}=0$. From the linear combinations of the equations (3.12) to (3.15) in Sec. 3.3 (and using (3.21)) we obtain the independent equations
\begin{equation}
\frac{d}{d {\hat r}}\bigg( N f \gamma R^5 \frac{d}{d {\hat r}} \ln \gamma^{\frac{1}{5}} R \bigg)  + \frac{1}{10} \, \frac{N}{f \gamma} \, e^{-M} \,\frac{Q_D^2}{R^5} - 4 Nf \gamma R^5 \, \frac{\alpha}{f^2 R^2} \nonumber
\end{equation}
\begin{equation}
 =\, - \frac{1}{5} \kappa^2_{10} g_s^2 \,\frac{N \gamma R^5}{f}\, {e^{\frac{2}{5} {\Phi_s} + \frac{6}{5} B}} \, T (\Phi) \delta^6 (\vec{\hat{r}}\,) \,\,,
\end{equation}
\begin{equation}
\frac{d}{d {\hat r}}\bigg( N f \gamma R^5 \frac{d}{d {\hat r}} \ln \gamma^{\frac{1}{5}} N \bigg)  - \frac{2}{5} \, \frac{N}{f \gamma} \, e^{-M} \,\frac{Q_D^2}{R^5} +6 Nf \gamma R^5 \, (\alpha - \alpha_{ij})\frac{1}{f^2 R^2} \nonumber
\end{equation}
\begin{equation}
 =\,\frac{4}{5} \kappa^2_{10} g_s^2 \,\frac{N \gamma R^5}{f}\, {e^{\frac{2}{5} {\Phi_s} + \frac{6}{5} B}} \, T (\Phi) \delta^6 (\vec{\hat{r}}\,) \,\,,
\end{equation}
\begin{equation}
\frac{d}{d{\hat r}}\bigg( N f \gamma R^5 \frac{d}{d {\hat r}} \ln \gamma^{-\frac{4}{5}} N \bigg)  - \frac{2}{5} \, \frac{N}{f \gamma} \, e^{-M} \,\frac{Q_D^2}{R^5} +2 Nf \gamma R^5 \, \Big(\sum_{i=1}^{2} \Big(\frac{p_i}{q_i}\Big)^2 \gamma^2 -2 \alpha \Big) \frac{1}{f^2 R^2} \nonumber
\end{equation}
\begin{equation}
 =\,\frac{4}{5} \kappa^2_{10} g_s^2\,\frac{N \gamma R^5}{f}\, {e^{\frac{2}{5} {\Phi_s} + \frac{6}{5} B}} \, T (\Phi) \delta^6 (\vec{\hat{r}}\,) \,\,.
\end{equation}
Further, from (5.1) and (5.2) we obtain
\begin{equation}
N f \gamma R^5 \frac{d}{d {\hat r}}\bigg(N f \gamma R^5 \frac{d}{d {\hat r}} \ln N \gamma R^4 \bigg) = (10\alpha +6 \alpha_{ij}) \big( N \gamma R^4 )^2  \,\,,
\end{equation}
and similarly from (5.2) and (5.3)
\begin{equation}
N f \gamma R^5 \frac{d}{d {\hat r}}\bigg(N f \gamma R^5 \frac{d}{d {\hat r}} \ln \gamma \bigg) = \Big( -10\alpha +6 \alpha_{ij} + 2\sum_{i=1}^{2} \Big(\frac{p_i}{q_i}\Big)^2 \gamma^2 \Big) \big( N \gamma R^4 )^2  \,\,.
\end{equation}

To solve the above equations we introduce a new variable $\rho$ defined by
\begin{equation}
N f \gamma R^5 \frac{d}{d {\hat r}} = \Big(\frac{\beta}{4} \Big) \rho^5 \frac{d}{d \rho} \,\,,
\end{equation}
where $\beta$ is defined by $\beta^2 = 10\alpha +6 \alpha_{ij}$. When $\alpha$ and $\alpha_{ij}$ are constants, which is indeed the case in our discussions in Secs. 5.1 and 5.2, (5.4) and (5.5) can be rewritten as
\begin{equation}
\rho^5 \frac{d}{d \rho} \bigg(\rho^5 \frac{d}{d \rho} \ln N \gamma R^4 \bigg) = 16 \big( N \gamma R^4 )^2  \,\,,
\end{equation}
and
\begin{equation}
\rho^5 \frac{d}{d \rho} \bigg(\rho^5 \frac{d}{d \rho} \ln \gamma \bigg) = \frac{16}{\beta^2} \Big(-10\alpha +6 \alpha_{ij} + 2\sum_{i=1}^{2} \Big(\frac{p_i}{q_i}\Big)^2 \gamma^2 \Big) \big( N \gamma R^4 )^2  \,\,.
\end{equation}
Among these equations (5.7) can be solved by
\begin{equation}
N \gamma R^4 = \rho^4 \,\,,
\end{equation}
and from (5.6) and (5.9) we have
\begin{equation}
\frac{d {\hat r}}{fR} = \Big(\frac{4}{\beta} \Big) \frac{d \rho}{\rho} \,\,.
\end{equation}
Now we solve the remaining equations for the two cases considered in Secs. 4.1 and 4.2.

\vskip 0.3cm
\hspace{-0.6cm}{\bf \large 5.1 Case I: $\gamma ({\hat r}) ={\rm constant} \equiv \gamma_0$}
\vskip 0.15cm

In this section we consider the case $\gamma ({\hat r}) ={\rm constant} \equiv \gamma_0$ as in Sec. 4.1. Since this section is an extension of Sec. 4.1, in the followings we basically consider the two cases in (4.14) and (4.15) for the definiteness of our discussion.  In the cases of (4.14) and (4.15), $\alpha$ and $\alpha_{ij}$ are equal to one, $\alpha =\alpha_{ij} =1$ (see (4.24) and (4.25)), and therefore $\beta =4$. Also using (4.20) one finds that (5.8) reduces to
\begin{equation}
\rho^5 \frac{d}{d \rho} \bigg( \rho^5 \frac{d}{d \rho} \ln \gamma \bigg) = 0 \,\,,
\end{equation}
while (5.7) remains unchanged. Since (5.7) remains unchanged, the solution is still given by (5.9). Also (5.11) is trivially satisfied by $\gamma ({\hat r}) ={\rm constant}$, so we can ignore them.

Now we go back to the original equations (5.1), (5.2) and (5.3). Using (5.6) with $\beta =4$ and (5.9), one can rewrite them as\footnote{Note that the third terms of (5.2) and (5.3) both vanish in the cases A and B in (4.14) and (4.15) and therefore (5.3) becomes identical to (5.2) in both of these cases.}
\begin{equation}
\nabla^2 \ln \gamma^{\frac{1}{5}}_0 R + \frac{1}{10} N^2 \, e^{-M} \,\frac{Q_D^2}{\rho^{10}} - \frac{4}{\rho^2} \,=\, - \frac{1}{5} \kappa^2_{10}  g_s^2 N\, {e^{\frac{2}{5} {\Phi_s} + \frac{6}{5} B}} \, T (\Phi) \delta^6 (\vec{\rho}\,) \,\,,
\end{equation}
\begin{equation}
\nabla^2 \ln \gamma^{\frac{1}{5}}_0 N - \frac{2}{5} N^2 \, e^{-M} \,\frac{Q_D^2}{\rho^{10}} \,=\, \frac{4}{5} \kappa^2_{10} g_s^2 N\, {e^{\frac{2}{5} {\Phi_s} + \frac{6}{5} B}} \, T (\Phi) \delta^6 (\vec{\rho}\,) \,\,,
\end{equation}
and similarly (3.19) and (3.20) as
\begin{equation}
\nabla^2 \Phi_s =0 \,\,,
\end{equation}
\begin{equation}
\nabla^2 B - \frac{1}{2} N^2 \, e^{-M} \,\frac{Q_D^2}{\rho^{10}} \,=\, \kappa^2_{10} g_s^2 N\, {e^{\frac{2}{5} {\Phi_s} + \frac{6}{5} B}} \, T (\Phi) \delta^6 (\vec{\rho}\,) \,\,,
\end{equation}
where the Laplacian $\nabla^2$ is defined by $\nabla^2 \equiv (1/\rho^5 )(d/d\rho)(\rho^5 d/d\rho)$, so the delta-function $\delta^6 (\vec{\rho}\,)$ is now normalized by $\int \rho^5 d\rho \epsilon_5 \delta^6 (\vec{\rho}\,)=1$. (This $\delta^6 (\vec{\rho}\,)$ is related to the original $\delta^6 (\vec{\hat{r}}\,)$ by the equation $\rho^6 \delta^6 (\vec{\rho}\,) =\Big(\frac{4}{\beta} \Big) \gamma R^6 \delta^6 (\vec{\hat{r}}\,)$.)

Now comparing (5.12) and (5.13) with (5.15) we obtain
\begin{equation}
N= \gamma^{-\frac{1}{5}}_0 e^{\frac{4}{5} B} \,\,,
\end{equation}
and
\begin{equation}
R= \gamma^{-\frac{1}{5}}_0 e^{-\frac{B}{5}} \rho \,\,,
\end{equation}
where $B(\rho)$ is a solution to the field equation (5.15), while $\Phi_s (\rho)$ is given by $\Phi_s (\rho) =0$. Finally from (5.10) and (5.17) we find that
\begin{equation}
\frac{d {\hat r}}{f} = \gamma^{-\frac{1}{5}}_0 e^{-\frac{B}{5}} d\rho \,\,,
\end{equation}
and therefore the 10D metric in (2.2) reduces to
\begin{equation}
ds_{10}^2 = e^B \Big( -dt^2 + d {\vec x}_3^2 \Big) + e^{-B} \Big( dr^2 + r^2 d \Sigma_{p_{_1}, p_{_2}}^{2} \Big) \,\,,
\end{equation}
where $t \equiv \gamma^{-\frac{1}{5}}_0 {\hat t}$ and $r \equiv \gamma^{-\frac{1}{5}}_0 \rho$. (5.19) is the usual background metric with D3-branes located at the conifold singularity (see eq. (5.10) of Ref.\cite{4}) and it reduces to (4.36) with ${\hat a}=0$ in the limit $Q_D  \rightarrow 0$.\footnote{The factor $e^B$ in (5.19) is given by $e^{B(r)} = \Big( 1+ \frac{Q_0}{r^{4}} \Big)^{-1/2}$ with $Q_0 = \frac{Q_D}{4}$ as in Ref.\cite{4}. So $e^B$ reduces $e^B \rightarrow 1$ in the limit $Q_D  \rightarrow 0$. See also Sec. 5.3 of this paper.}

\vskip 0.3cm
\hspace{-0.6cm}{\bf \large 5.2 Case II: $\gamma ({\hat r}) = \frac{b}{\hat r}$ with $p_1 =p_2 =0$ }
\vskip 0.15cm

In Sec. 5.1 we obtained the usual D3-brane solution by taking $\gamma ({\hat r}) =  {\rm constant}$. In this section we will consider the case where $\gamma ({\hat r})$ and $p_i$ are given by $\gamma ({\hat r}) = \frac{b}{\hat r}$ and $p_1 =p_2 \equiv p =0$ as in Sec. 4.2 (see (4.47)). In this case $\alpha$ and $\alpha_{ij}$ are constants :
\begin{equation}
\alpha = \frac{6}{5q}\,\,,~\,\alpha_{ij}=\frac{1}{q}\,\,~~~\rightarrow~~~\beta^2 = \frac{18}{q}\,\,,
\end{equation}
(see (4.38)) and (5.8) reduces to
\begin{equation}
\rho^5 \frac{d}{d \rho} \bigg( \rho^5 \frac{d}{d \rho} \ln \gamma \bigg) = -\frac{16}{3} \big( N \gamma R^4 )^2  \,\,.
\end{equation}
Using (5.9) one finds that (5.21) is solved by
\begin{equation}
\gamma = \Big( \frac{\rho}{\rho_0} \Big)^{-\frac{4}{3}} \,\,.
\end{equation}

Now we go back to (5.1), (5.2) and (5.3). Using (5.6) and (5.9) we can rewrite them as
\begin{equation}
\nabla^2 \ln \gamma^{\frac{1}{5}} R + \frac{1}{10} \Big(\frac{4}{\beta} \Big)^2 N^2 \, e^{-M} \,\frac{Q_D^2}{\rho^{10}} - 4 \alpha \Big(\frac{4}{\beta} \Big)^2 \frac{1}{\rho^2} \,=\, - \frac{1}{5} \kappa^2_{10} g_s^2 \Big(\frac{4}{\beta} \Big) N\, {e^{\frac{2}{5} {\Phi_s} + \frac{6}{5} B}} \, T (\Phi) \delta^6 (\vec{\rho}\,) \,\,,
\end{equation}
\begin{equation}
\nabla^2 \ln \gamma^{\frac{1}{5}} N - \frac{2}{5} \Big(\frac{4}{\beta} \Big)^2 N^2 \, e^{-M} \,\frac{Q_D^2}{\rho^{10}} + 6 (\alpha - \alpha_{ij}) \Big(\frac{4}{\beta} \Big)^2 \frac{1}{\rho^2} \,=\, \frac{4}{5} \kappa^2_{10} g_s^2 \Big(\frac{4}{\beta} \Big) N\, {e^{\frac{2}{5} {\Phi_s} + \frac{6}{5} B}} \, T (\Phi) \delta^6 (\vec{\rho}\,) \,\,,
\end{equation}
\begin{equation}
\nabla^2 \ln \gamma^{-\frac{4}{5}} N - \frac{2}{5} \Big(\frac{4}{\beta} \Big)^2 N^2 \, e^{-M} \,\frac{Q_D^2}{\rho^{10}} + 2 \Big(\sum_{i=1}^{2} \Big(\frac{p_i}{q_i}\Big)^2 \gamma^2 -2 \alpha \Big) \Big(\frac{4}{\beta} \Big)^2 \frac{1}{\rho^2} \nonumber
\end{equation}
\begin{equation}
=\, \frac{4}{5} \kappa^2_{10} g_s^2 \Big(\frac{4}{\beta} \Big) N\, {e^{\frac{2}{5} {\Phi_s} + \frac{6}{5} B}} \, T (\Phi) \delta^6 (\vec{\rho}\,) \,\,,
\end{equation}
while the equations for $\Phi_s$ and $B$ are given by $\nabla^2 \Phi_s =0$ and
\begin{equation}
\nabla^2 B - \frac{1}{2} \Big(\frac{4}{\beta} \Big)^2 N^2 \, e^{-M} \,\frac{Q_D^2}{\rho^{10}} \,=\, \kappa^2_{10} g_s^2 \Big(\frac{4}{\beta} \Big) N\, {e^{\frac{2}{5} {\Phi_s} + \frac{6}{5} B}} \, T (\Phi) \delta^6 (\vec{\rho}\,) \,\,,
\end{equation}
where $\nabla^2$ is defined by $\nabla^2 \equiv ({1}/{\rho^5}) ({d}/{d\rho}) ({\rho^5 d}/{d \rho})$ as before, and $\alpha$, $\alpha_{ij}$ and $\beta$ are given by (5.20). Now the linear combinations of (5.24) and (5.25) give
\begin{equation}
\nabla^2 \ln N^5 -2 \Big(\frac{4}{\beta} \Big)^2 N^2 \, e^{-M} \,\frac{Q_D^2}{\rho^{10}} \,=\, 4\kappa^2_{10} g_s^2 \Big(\frac{4}{\beta} \Big) N\, {e^{\frac{2}{5} {\Phi_s} + \frac{6}{5} B}} \, T (\Phi) \delta^6 (\vec{\rho}\,) \,\,,
\end{equation}
\begin{equation}
\nabla^2 \ln \gamma + \frac{6}{q} \Big(\frac{4}{\beta} \Big)^2 \frac{1}{\rho^2} \,=\,0 \,\,,
\end{equation}
and from (5.23) and (5.28) one obtains
\begin{equation}
\nabla^2 \ln R^5 + \frac{1}{2} \Big(\frac{4}{\beta} \Big)^2 N^2 \, e^{-M} \,\frac{Q_D^2}{\rho^{10}} -  \frac{30}{q} \Big(\frac{4}{\beta} \Big)^2 \frac{1}{\rho^2} \,=\, -\kappa^2_{10} g_s^2 \Big(\frac{4}{\beta} \Big) N\, {e^{\frac{2}{5} {\Phi_s} + \frac{6}{5} B}} \, T (\Phi) \delta^6 (\vec{\rho}\,) \,\,.
\end{equation}

In the above equations (5.28) is consistent with (5.22) because $\frac{6}{q} (\frac{4}{\beta})^2 =  \frac{16}{3}$. So we can ignore it and we only need to solve (5.27) and (5.29). Comparing (5.27) with (5.26) one obtains
\begin{equation}
N= e^{\frac{4}{5} B} \,\,.
\end{equation}
Similarly, comparing (5.29) with (5.26) one finds that $R$ must be of the form
\begin{equation}
R= e^{-\frac{B}{5}} {\hat R} \,\,,
\end{equation}
where ${\hat R}$ is defined by the equation
\begin{equation}
\nabla^2 \ln {\hat R} - \frac{6}{q} \Big(\frac{4}{\beta} \Big)^2 \frac{1}{\rho^2} \,=\,0 \,\,.
\end{equation}
Then using (5.22), (5.30) and (5.31) one finds from (5.9) that
\begin{equation}
{\hat R} = \Big( \frac{\rho}{\rho_0} \Big)^{\frac{1}{3}} \rho \,\,,
\end{equation}
which is consistent with (5.28) and (5.32) because those two equations imply $\gamma {\hat R} = {\rm constant}$, and this is indeed the case as one can check from (5.22) and (5.33). We have
\begin{equation}
\gamma {\hat R} = \rho_0 \,\,.
\end{equation}

Finally from (5.10) and (5.31) with (5.33) we find that
\begin{equation}
\frac{d {\hat r}}{f} = \Big(\frac{4}{\beta} \Big) e^{-\frac{B}{5}} \Big( \frac{\rho}{\rho_0} \Big)^{\frac{1}{3}} d{\rho} \,\,.
\end{equation}
and collecting all these together we obtain
\begin{equation}
ds_{10}^2 = e^B \Big( -dt^2 + d {\vec x}_3^2 \, \Big) + e^{-B} \Bigg( \Big(\frac{4}{\beta} \Big)^2 \Big( \frac{\rho}{\rho_0} \Big)^{\frac{2}{3}} d\rho^2 + \frac{\rho_0^2}{9} d \psi^2 \nonumber
\end{equation}
\begin{equation}
+ \Big( \frac{\rho}{\rho_0} \Big)^{\frac{2}{3}} \rho^2 \sum_{i=1}^{2} \frac{q_i}{6} \Big( d\theta_i^2 + \sin^2 \theta_i d\phi_i^2 \,\Big) \Bigg) \,\,,
\end{equation}
where we have changed $d{\hat t} \,\rightarrow \, dt$. Now we finally introduce a new variable $r$ defined by
\begin{equation}
r = \Big( \frac{3}{\beta} \Big) \, \rho\, \Big( \frac{\rho}{\rho_0} \Big)^{\frac{1}{3}}  \,\,.
\end{equation}
Then we obtain the final version of the 10D metric
\begin{equation}
ds_{10}^2 = e^B \Big( -dt^2 + d {\vec x}_3^2 \, \Big) + e^{-B} \Bigg( dr^2 + \frac{r^2}{3} \sum_{i=1}^{2} \Big( d\theta_i^2 + \sin^2 \theta_i d\phi_i^2 \,\Big) + R_0^2  d \psi^2 \Bigg) \,\,,
\end{equation}
where $R_0$ is given by $R_0 = \frac{\rho_0}{3}$.

The metric (5.38) is unique in the sense that it does not involve NS5-branes even under T-duality
as we shall see in Sec. VI. Namely (5.38) and its T-dual partner describe torsion-free background configurations in which the two-form field $B_{(2)}$ is absent. The metric (5.38) describes a background geometry with D3-branes located at the singularity $r=0$ of the cone-type internal space whose topology at constant $r$ is given by $S_2 \times S_2 \times S_1$, while its T-dual partner describes D4-branes located at $r=0$ of the internal space with topology $S_2 \times S_2$ (see Sec. VI). (5.38) reduces to (4.49) with $q=2$ and ${\hat a}=0$ in the absence of D3-branes because $e^B$ in (5.38) becomes $e^B \rightarrow 1$ when the D-brane charge $Q_D$ vanishes (see Sec. 5.3).

\vskip 0.3cm
\hspace{-0.6cm}{\bf \large 5.3 The factor $e^B$ of the torsion-free background metric}
\vskip 0.15cm

In Sec. 5.2 we obtained (5.38), a torsion-free background solution in the presence of D3-branes. To obtain (5.38) we have used three equations (5.1) to (5.3). But these three equations were originally obtained from (3.12) to (3.15) by eliminating $\mathcal{H}$ in three different ways. Since equations (5.1) to (5.3) are three equations, whereas (3.12) to (3.15) are four, we need to consider one more independent equation which now contains $\mathcal{H}$, and be consistent with the given solution (5.38) of Sec. 5.2. The simplest such an equation may be obtained by subtracting (3.13) from (3.12). We have
\begin{equation}
\frac{1}{R} f \big( f R^{\prime})^{\prime} - f^2 \frac{N^{\prime}}{N}\frac{R^{\prime}}{R} + \frac{2}{25} f^2 \mathcal{H} + \frac{1}{5} \frac{1}{\gamma}f \big( f \gamma^{\prime})^{\prime} + \frac{2}{5} f^2 \frac{R^{\prime}}{R}\frac{\gamma^{\prime}}{\gamma} - \frac{1}{5} f^2 \frac{N^{\prime}}{N}\frac{\gamma^{\prime}}{\gamma} \nonumber
\end{equation}
\begin{equation}
= - \frac{1}{5} \kappa_{10}^2 g_s^2   e^{\frac{2}{5} {\Phi_s} + \frac{6}{5} B}  T (\Phi) \delta^6 ({\vec{\hat r}}\,) \,\,,
\end{equation}
where the "prime" denotes $\prime \equiv \frac{d}{d{\hat{r}}}$ as before.

To see what (5.39) means we first change the variable $\hat{r}$ in (5.39) into $r$ as follows. From (5.37) we have
\begin{equation}
dr = \Big( \frac{4}{\beta} \Big) \, \Big( \frac{\rho}{\rho_0} \Big)^{\frac{1}{3}} d \rho  \,\,,
\end{equation}
and from (5.35) and (5.40),
\begin{equation}
\frac{d\hat{r}}{f} = e^{-\frac{B}{5}} dr ~~~\rightarrow~~~f \frac{d}{d\hat{r}} = e^{\frac{B}{5}} \frac{d}{dr} \,\,.
\end{equation}
So using (5.41), one can rewrite (5.39) into
\begin{equation}
 e^{-\frac{B}{5}} \frac{1}{R} \frac{d}{dr} \Big( e^{\frac{B}{5}} \frac{dR}{dr} \Big) - \Big( \frac{1}{N} \frac{dN}{dr}\Big)\Big( \frac{1}{R} \frac{dR}{dr}\Big) + \frac{6}{25} \Big(\frac{dB}{dr}\Big)^2 + \frac{1}{5} e^{-\frac{B}{5}}  \frac{1}{\gamma} \frac{d}{dr} \Big( e^{\frac{B}{5}} \frac{d\gamma}{dr} \Big) \nonumber
\end{equation}
\begin{equation}
+ \frac{2}{5}\Big( \frac{1}{R} \frac{dR}{dr}\Big)\Big( \frac{1}{\gamma} \frac{d\gamma}{dr} \Big) - \frac{1}{5}\Big( \frac{1}{N} \frac{dN}{dr}\Big)\Big( \frac{1}{\gamma} \frac{d\gamma}{dr} \Big)= - \frac{1}{5} \kappa_{10}^2 g_s^2  e^{\frac{4}{5} B}  T (\Phi) \delta^6 ({\vec{\hat r}}\,) \,\,,
\end{equation}
where, and in what follows, we will set $\Phi_s =0$.

In the next we change the derivatives $\frac{dR}{dr}$ and $\frac{dN}{dr}$ in (5.42) into $\frac{dB}{dr}$ by using (5.30) and (5.31). From (5.31) together with (5.33) and (5.37), we have
\begin{equation}
R(r) = \Big( \frac{\beta}{3} \Big) e^{-\frac{B}{5}} r \,\,,
\end{equation}
and therefore
\begin{equation}
\frac{dR}{dr} = \Big( \frac{\beta}{3} \Big) e^{-\frac{B}{5}} \Big(1- \frac{r}{5} \frac{dB}{dr} \Big)\,\,.
\end{equation}
Also from (5.30), $\frac{dN}{dr}$ becomes
\begin{equation}
\frac{dN}{dr} = \frac{4}{5} e^{\frac{4}{5}B} \Big(\frac{dB}{dr} \Big)\,\,,
\end{equation}
and from (5.22) and (5.37) we have
\begin{equation}
\gamma = \Big( \frac{3}{\beta} \Big) \frac{\rho_0}{r}~~~\rightarrow~~~\frac{d\gamma}{dr}=-\Big( \frac{3}{\beta} \Big) \frac{\rho_0}{r^2} \,\,.
\end{equation}
Substituting these equations into (5.42), we obtain
\begin{equation}
\frac{d^2 B}{dr^2} + \frac{4}{r} \frac{dB}{dr} -2\Big(\frac{dB}{dr} \Big)^2 = \kappa_{10}^2 g_s^2 e^{\frac{4}{5} B}  T (\Phi) \delta^6 ({\vec{\hat r}}\,) \,\,.
\end{equation}

Now we finally change the variable $r$ in (5.47) into $\rho$. From (5.37) we have
\begin{equation}
\frac{d}{dr} = \Big( \frac{\beta}{4} \Big) \, \Big( \frac{\rho}{\rho_0} \Big)^{-\frac{1}{3}} \frac{d}{d \rho}  \,\,,
\end{equation}
and using $\delta^6 ({\vec{\hat r}}) =\big( \frac{\beta}{4} \big) \frac{\rho^6}{\gamma R^6} \delta^6 ({\vec{\rho}})$ together with (5.37) and (5.48), we can change (5.47) into
\begin{equation}
\nabla^2 B - 2 \Big(\frac{dB}{d\rho} \Big)^2 = \kappa_{10}^2 g_s^2 \Big( \frac{4}{\beta} \Big) e^{2B} T (\Phi) \delta^6 ({\vec{\rho}}) \,\,,
\end{equation}
where $\nabla^2 \equiv (1/\rho^5 )(d/d\rho)(\rho^5 d/d\rho)$. (5.49) is the final version of (5.39) and it is another second order equation for $B$ besides (5.26). Note that (5.26) can be rewritten as
\begin{equation}
\nabla^2 B - \frac{1}{2} \Big( \frac{4}{\beta} \Big)^2 e^{4B} \frac{Q_D^2}{\rho^{10}} = \kappa_{10}^2 g_s^2 \Big( \frac{4}{\beta} \Big) e^{2B} T (\Phi) \delta^6 ({\vec{\rho}}) \,\,,
\end{equation}
if we use (5.30) and $\Phi_s =0$. If we define
\begin{equation}
e^{4B} \equiv \chi \,\,,
\end{equation}
(5.49) and (5.50) can be rewritten as
\begin{equation}
\nabla^2 \ln \chi - \frac{1}{2} \Big( \frac{d}{d\rho} \ln \chi \Big)^2 = 2 c_B \delta^6 ({\vec{\rho}}) \,\,,
\end{equation}
\begin{equation}
\nabla^2 \ln \chi - 2 \frac{q_0^2}{\rho^{10}} \chi = 2 c_B \delta^6 ({\vec{\rho}}) \,\,,
\end{equation}
where $q_0 \equiv ( \frac{4}{\beta}) Q_D$ and the constant $c_B$ is defined by
\begin{equation}
c_B \equiv 2 \kappa_{10}^2 g_s \Big( \frac{4}{\beta} \Big) \chi^{1/2} (0) T_3 \,\,.
\end{equation}

Now subtracting (5.52) from (5.53) we obtain
\begin{equation}
\Big( \frac{d}{d\rho} \ln \chi \Big)^2 - 4 \frac{q_0^2}{\rho^{10}} \chi = 0 \,\,,
\end{equation}
and one observes that (5.53) and (5.55) precisely coincide with the equations (5.3) and (5.4) of Ref.\cite{4} only except that the variable $r$ in Ref.\cite{4} is replaced by $\rho$ in this paper. So the solution to the equations (5.52) and (5.53) must take the same form as (5.9) of Ref.\cite{4}. We have
\begin{equation}
e^{B(\rho)} =  \Big( 1+ \frac{Q_0}{\rho^{4}} \Big)^{-1/2}\,\,,~~~~~\Big( Q_0 \equiv \frac{q_0}{4} \Big)\,\,.
\end{equation}
Further, if we use (5.37), we obtain
\begin{equation}
e^{B(r)} =  \Big( 1+ \frac{r_0^3}{r^{3}} \Big)^{-1/2}\,\,,
\end{equation}
where $r_0 \equiv \big( \frac{3}{\beta} \big)\big(\frac{Q_0}{\rho_0} \big)^{1/3}$.
This is the final form of $e^{B(r)}$ in (5.38). Note that this $e^{B(r)}$ reduces $e^B \rightarrow 1$ when $Q_D \rightarrow 0$. Namlely the metric (5.38) reduces to (4.49) with $q=2$ and ${\hat a}=0$ when D-branes are absent.

\vskip 1cm
\hspace{-0.65cm}{\bf \Large VI. Summary and discussions}
\vskip 0.5cm
\setcounter{equation}{0}
\renewcommand{\theequation}{6.\arabic{equation}}

In this paper we obtained a Ricci-flat solution describing a background vacuum of the string theory. This solution has an exceptional property distinguished from the usual background solutions. It does not involve NS5-branes under T-duality unlikely to the case of the ordinary Calabi-Yau (or the conifold) ansatz. As mentioned in the introduction a conifold singularity of the Calabi-Yau ansatz becomes two intersecting NS5-branes under T-duality transformation. So the IIB metric of D3-brane at a conifold singularity corresponds to a IIA configuration of D4-brane suspended between two orthogonal NS5-branes.\cite{8} The background solution (5.38), however, does not have this property.

Under T-duality (5.38) becomes
\begin{equation}
ds_{10 {\rm T}}^{2} = e^B \Big( -dt^2 + d {\vec x}_3^2 + {\tilde R}_0^2 d \psi^2 \Big) + e^{-B} \Big( dr + \frac{r^2}{3} \sum_{i=1}^{2} \big( d\theta_i^2 + \sin^2 \theta_i d\phi_i^2 \big) \Big) \,\,,
\end{equation}
where ${\tilde R}_0$ is defined by ${\tilde R}_0 \equiv \frac{1}{R_0}$ in the unit $\alpha^{\prime} =1$. The metric (6.1) describes a (stack of) D4-brane(s) located at the singularity $r=0$ of the cone-type internal space whose topology at constant $r$ is given by $S_2 \times S_2$. But in this configuration NS5-branes are absent because the NS-NS two-form fields are zero, though the D4-branes of the IIA configuration is still present. Note that under T-duality the 6D metric $h_{ab}$ and the two-form field $B_{ab}$ transform as \cite{9}
\begin{equation}
\tilde{h}_{\psi\psi}  =  \frac{1}{h_{\psi\psi}}\,\, , ~~~~~~\,\,~ e^{2\tilde{\Phi}}  =   \frac{e^{2\Phi}}{h_{\psi\psi}} \,\,,~~~~~~~~~~~~~~~~~~~~~~ \nonumber
\end{equation}
\begin{equation}
\tilde{h}_{a\psi}  =  \frac{B_{a\psi}}{h_{\psi\psi}} \,\,,~~~~~~\,\,~
\tilde{B}_{a\psi} = \frac{h_{a\psi}}{h_{\psi\psi}} \,\,,~~~~~~~~(a,b \neq \psi)  \,\,,\nonumber
\end{equation}

\begin{equation}
\tilde{h}_{ab}  =  h_{ab}-\frac{h_{a\psi}h_{b\psi} - B_{a\psi} B_{b\psi}}{h_{\psi\psi}} \,\, ,~~~~~~~~~~~~~~~~~~~~~~ \nonumber
\end{equation}

\begin{equation}
\tilde{B}_{ab}  =  B_{ab}-\frac{B_{a\psi}h_{b\psi} - B_{b\psi} h_{a\psi}}{h_{\psi\psi}} \,\, ,~~~~~~~~~~~~~~~~~~~~~~~~
\end{equation}
where ${B}_{ab} = {B}_{a\psi} =0$ in our case. From (6.2) one finds that the two-form fields $\tilde{B}_{ab}$ and $\tilde{B}_{a\psi}$ all vanish because the only nonvanishing components $\tilde{B}_{\phi_i \psi}$ are given by $\tilde{B}_{\phi_i \psi}  \propto p_i \cos \theta_i$, but $p_i =0$ in our metric (5.38). So the NS-NS two-form fields of the T-dual configuration all vanish and the NS5-branes are absent in this configuration.

Indeed, $p_i =0$ is a necessary condition for a 6D metric to be Ricci-flat when $h_{\psi\psi}$ is constant. To see this, note that the Ricci-tensors of the 6D metric
\begin{equation}
ds_{6}^{2} = \frac{dr^2}{f^2 (r)} + \frac{U^2 (r)}{9} \Big( d \psi + \sum_{i=1}^{2} p_i \cos \theta_i d \phi_i \Big)^2 + \frac{V^2 (r)}{6} \sum_{i=1}^{2} q_i \Big( d\theta_i^2 + \sin^2 \theta_i d\phi_i^2 \Big) \,\,
\end{equation}
are given (in the orthonormal basis) by
\begin{equation}
{\mathcal R}_{rr} = -\big(U^{\prime} f\,\big)^{\prime} \frac{f}{U} -4 \big(V^{\prime} f\,\big)^{\prime} \frac{f}{V} \,\,,
\end{equation}
\begin{equation}
{\mathcal R}_{\theta_i \theta_i} (={\mathcal R}_{\phi_i \phi_i} ) = - \big(V^{\prime} f\,\big)^{\prime} \frac{f}{V} - f^2 \Big(3\frac{V^{\prime}}{V} + \frac{U^{\prime}}{U} \Big)\frac{V^{\prime}}{V} -2 \Big( \frac{p_i}{q_i} \Big)^2 \Big(\frac{U}{V^2} \Big)^2 + \frac{6}{q_i} \frac{1}{V^2} \,\,,
\end{equation}
\begin{equation}
{\mathcal R}_{\psi\psi} = -\big(U^{\prime} f\,\big)^{\prime} \frac{f}{U} -4 f^2 \frac{U^{\prime}}{U} \frac{V^{\prime}}{V} +2 \sum_{i=1}^{2} \Big( \frac{p_i}{q_i} \Big)^2 \Big(\frac{U}{V^2} \Big)^2  \,\,.
\end{equation}
From (6.5) one finds that ${\mathcal R}_{\psi\psi}$ vanishes only for $p_i =0$ when $U$ is a constant.

In addition to this, one can check that ${\mathcal R}_{rr}$ and ${\mathcal R}_{\theta_i \theta_i}$($={\mathcal R}_{\phi_i \phi_i}$) also vanish for $dS_6^2$ in (5.38), as well as ${\mathcal R}_{\psi\psi}$. Namely (5.38) is a Ricci-flat metric which does not produce the NS-NS two-form fields under T-duality and therefore the T-dual partner of (5.38) dose not involve the NS-NS type branes in its background configuration. So the non-linear $\sigma$-models whose target space metrics are given by these T-dual partners can both be torsion-free.\cite{10}
\vskip 1cm
\begin{center}
{\large \bf Acknowledgement}
\end{center}

This research was supported by Basic Science Research Program through the National Research Foundation of Korea(NRF) funded by the Ministry of Education(Grant No. 2018R1D1A1B07050146).

\end{document}